\documentclass[12pt,onecolumn]{IEEEtran}
\usepackage[left=1.25in, right=1.25in, bottom=1in, top=1in]{geometry}

\usepackage{cite}
\usepackage{amsmath,amssymb,amsfonts}
\usepackage{algorithmic}
\usepackage{graphicx,color}
\usepackage{textcomp}
\usepackage{xcolor}
\usepackage{hyperref}
\hypersetup{hidelinks=true}
\usepackage{algorithm,algorithmic}
\usepackage{siunitx}
\usepackage{MnSymbol}
\usepackage{placeins}
\usepackage{subfigure}
\usepackage{float}

\allowdisplaybreaks

\renewcommand{\Vec}{\mathbf}

\begin{document}

\title{Identifiability Conditions for Acoustic Feedback Cancellation with the Two-Channel Adaptive Feedback Canceller Algorithm}\author{\thanks{© 2025 IEEE\\This research was carried out at the ESAT Laboratory of KU Leuven, in the frame of Research Council KU Leuven C14-21-0075 ”A holistic approach to the design of integrated and distributed digital signal processing algorithms for audio and speech communication devices”, and Aspirant Grant 11PDH24N (for A. Roebben) from the Research Foundation - Flanders (FWO). The scientific responsibility is assumed by its authors.}Arnout Roebben\thanks{Arnout Roebben, Toon van Waterschoot and Marc Moonen are with the Department of Electrical Engineering (ESAT), STADIUS Center for Dynamical Systems, Signal Processing and Data Analytics, KU Leuven, B-3001 Leuven, Belgium (e-mail: arnout.roebben@esat.kuleuven.be).}, Toon van Waterschoot, Jan Wouters,\thanks{Jan Wouters is with the Department of Neurosciences, Research Group ExpORL, KU Leuven, B-3000 Leuven, Belgium.} and Marc Moonen}

\maketitle

\begin{abstract}
	\IEEEPARstart{I}{n} audio signal processing applications with a microphone and a loudspeaker within the same acoustic environment, the loudspeaker signals can feed back into the microphone, thereby creating a closed-loop system that potentially leads to system instability. To remove this acoustic coupling, prediction error method (PEM) feedback cancellation algorithms aim to identify the feedback path between the loudspeaker and the microphone by assuming that the input signal can be modelled by means of an autoregressive (AR) model. It has previously been shown that this PEM framework and resulting algorithms can identify the feedback path correctly in cases where the forward path from microphone to loudspeaker is sufficiently time-varying or non-linear, or when the forward path delay equals or exceeds the order of the AR model. In this paper, it is shown that this delay-based condition can be generalised for one particular PEM-based algorithm, the so-called two-channel adaptive feedback canceller (2ch-AFC), to an invertibility-based condition, for which it is shown that identifiability can be achieved when the order of the forward path feedforward filter exceeds the order of the AR model. Additionally, the condition number of inversion of the correlation matrix as used in the 2ch-AFC algorithm can serve as a measure for monitoring the identifiability.
\end{abstract}

\begin{IEEEkeywords}
    Acoustic feedback cancellation, Adaptive feedback cancellation, Identifiability conditions, Two-channel adaptive feedback canceller
\end{IEEEkeywords}

\section{INTRODUCTION} \label{sec:introduction}
\begin{figure}
	\centering
	\includegraphics[width=0.5\linewidth]{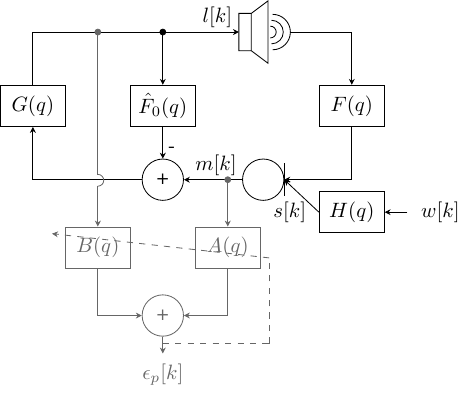}
	\caption{Identification of feedback path $F(q)$ in a closed-loop microphone-loudspeaker system using the two-channel adaptive feedback canceller (2ch-AFC) algorithm\cite{spriet_adaptive_2005}.}
	\label{fig:Signal_model}
\end{figure}

\IEEEPARstart{I}{n} audio signal processing applications with a microphone and a loudspeaker within the same acoustic environment, e.g., as in acoustic hearing instruments or public address (PA) systems, the loudspeaker signals can feed back into the microphone, creating a closed-loop system \cite{dillon_hearing_2012,sankowsky-rothe_acoustic_2020,spriet_evaluation_2010,van_waterschoot_fifty_2011,kates_digital_2008}. This closed-loop system, as shown in Fig. \ref{fig:Signal_model}, can be rendered unstable for amplifications in the forward path ($G(q)$ in Fig. 
\ref{fig:Signal_model}) exceeding the maximum stable gain (MSG), thereby resulting in undesired acoustic howling \cite{van_waterschoot_fifty_2011}. This is problematic in, e.g., acoustic hearing instruments that rely on amplification to combat the user's hearing loss \cite{van_waterschoot_fifty_2011,dillon_hearing_2012,spriet_evaluation_2010}.

There exist various strategies to remove this acoustic howling and system instability, including reducing the gain only for the unstable frequencies by using a notch filter \cite{li_howling_2021,mikami_adaptive_2018}, modifying the phase \cite{guo_use_nodate} or frequency \cite{on_the_periodically_guo_2016,berdhal_frequency_2010} of the microphone signals, applying a (spatial) filter or neural network (NN) to the microphone signals \cite{rombouts_generalized_2006,zheng_deep_2022,zhang_deep_2023}, and adaptive feedback cancellation (AFC) \cite{van_waterschoot_fifty_2011,spriet_adaptive_2005,hellgren_bias_2001,hellgren_analysis_2002}. In this latter AFC strategy, loudspeaker and microphone signals are observed to identify a feedback path model using adaptive filters \cite{spriet_adaptive_2005,rombouts_acoustic_2006,boukis_toward_2007,vanitha_devi_curvelet_2023,guo_analysis_2011,guo_novel_2012,spriet_evaluation_2010}, possibly guided by NNs \cite{lydaki_deep_feedback_2025,soleimani_neural-afc_2023,zhang_neural_2023,deeppem_afc_zhan_2025}. The feedback path model can then be used to counteract the true acoustic feedback path. AFC algorithms have the advantage of not altering the input signal unlike using gain reduction, phase or frequency modification. 

However, using standard adaptive filters for AFC as those designed for acoustic echo cancellation (AEC) \cite{benesty_advances_2001,valero_multi-microphone_2017}, causes a bias in the identified feedback path due to the correlation between microphone and loudspeaker signals through the closed-loop \cite{guo_analysis_2011,van_waterschoot_fifty_2011}. Prediction error method (PEM) based (adaptive) identification resolves this bias-problem through assuming that the input signal adheres to an autoregressive (AR) model, such that the feedback path can be identified, jointly with this AR model \cite{hellgren_bias_2001}.

For this PEM-based identification to correctly identify the feedback path, identifiability conditions need to be satisfied. These identifiability conditions then lead to unbiased feedback path estimation that can be leveraged both by regular adaptive filters \cite{spriet_adaptive_2005,rombouts_acoustic_2006,boukis_toward_2007}, and NN-guided filters that reuse the same structure \cite{soleimani_neural-afc_2023,deeppem_afc_zhan_2025}. Additionally, as AFC algorithms are often applied in conjunction with NN-based filtering of the microphone signals, unbiased feedback path estimation is also desired in such cases \cite{zheng_deep_2022}. Therefore, in \cite{hellgren_bias_2001,spriet_adaptive_2005}, it is shown that identifiability can be obtained if the forward path from microphone to loudspeaker is sufficiently time-varying or non-linear, which is, e.g., leveraged in \cite{boukis_toward_2007}. Indeed, in \cite{boukis_toward_2007} time-varying all-pass infinite impulse response (IIR) filters are used to insert time-variation into the forward path. In \cite{spriet_adaptive_2005}, it is additionally shown that identifiability can be obtained if the forward path delay is equal to or exceeds the order of the AR model.  

Several algorithms have been developed for PEM-based identification, e.g., depending on whether the feedback path is directly estimated or estimated through an auxiliary filter \cite{spriet_adaptive_2005,rombouts_acoustic_2006}. One of these algorithms is the two-channel adaptive feedback canceller (2ch-AFC) \cite{spriet_adaptive_2005}, where the AR model is jointly estimated with an auxiliary filter, chosen such that the optimisation problem becomes linear-in-the-parameters. 

In this paper, this 2ch-AFC algorithm is considered. It will be shown that specifically for this 2ch-AFC algorithm, the delay-based identifiability condition requiring that the delay in the forward path is equal to or exceeds the order of the AR model $L_A$ \cite{spriet_adaptive_2005}, can be generalised to an invertibility-based condition requiring that the order of the forward path feedforward filter exceeds $L_A$. Thus, rather than requiring that the first $L_A$ coefficients of the forward path feedforward filter must be zero as in the delay-based condition, these coefficients can be non-zero, thereby allowing for a more flexible design of the signal processing in the forward path of hearing instruments and PA systems. The delay-based condition can also be seen as a special case of the latter condition. Indeed, when the first $L_A$ coefficients of the forward path feedforward filter are zero, both conditions align. Further, rather than requiring time-varying (all-pass) filters to obtain identifiability as in \cite{boukis_toward_2007}, identifiability can be obtained using stationary (all-pass) filters. It is also noted that this invertibility-condition is specifically derived for the 2ch-AFC algorithm, and does not immediately apply to other PEM-based feedback cancellation algorithms. Finally, it will be shown that the condition number of inversion of the correlation matrix used in the 2ch-AFC algorithm can serve as an identifiability measure. 

The signal model is introduced in Section \ref{sec:signal_model}, and the 2ch-AFC algorithm is reviewed in Section \ref{sec:two_channel_adaptive_feedback_canceller}. In Section \ref{sec:identifiability} the delay-based condition \cite{spriet_adaptive_2005} is revisited, and the invertibility-based condition is presented. This condition is verified in Section \ref{sec:results_and_discussion} according to the procedures as described in Section \ref{sec:experiment_design}. The code used for the experiments is available at \cite{roebben_github}. Finally, Section \ref{sec:conclusion} concludes this paper.  

\section{SIGNAL MODEL} \label{sec:signal_model}
The closed-loop microphone-loudspeaker system is shown in Fig. \ref{fig:Signal_model}. The finite impulse response (FIR) feedback path \cite{spriet_adaptive_2005}
\begin{equation}
F(q)=f[0]+f[1] q^{-1} + \cdots + f[L_F-1]q^{-L_F+1}_{\textstyle \raisebox{2pt}{,}}
\end{equation}
of order $L_F$ represents the acoustic coupling between the loudspeaker and microphone, and $q^{-1}$ represents the delay operator \cite{spriet_adaptive_2005}. The microphone signal $m[k]\in\mathbb{R}$ is given as
\begin{equation} \label{eq:m}
	m[k] = F(q)l[k] + s[k]_{\textstyle \raisebox{2pt}{.}}
\end{equation}
Herein, $k$ is the time-index, $F(q)l[k]$ represents the filtering of loudspeaker signal $l[k]$ with $F(q)$, and $s[k]$ represents the input signal impinging on the microphone \cite{spriet_adaptive_2005}. The filtering $F(q)l[k]$ can alternatively be represented in vector-notation as $F(q)l[k]=\Vec{f}^\top \Vec{l}[k]$ with
\begin{subequations}
	\begin{align} 
		\Vec{f} &= \begin{bmatrix}f[0] & \cdots & f[L_F-1]\end{bmatrix}^\top&\in\mathbb{R}^{L_F\times1}\\
		\Vec{l}[k] &= \begin{bmatrix}l[k] & \cdots & l[k-L_F+1]\end{bmatrix}^\top&\in\mathbb{R}^{L_F\times1}_{\textstyle \raisebox{2pt}{.}}
	\end{align}
\end{subequations}
Within the PEM framework \cite{spriet_adaptive_2005,hellgren_bias_2001}, this input signal $s[k]$ is furthermore assumed to be generated by passing white noise $w[k]\in\mathbb{R}$ through an AR model $H(q)$ as 
\begin{subequations}
	\begin{align} 
		s[k] &= H(q)w[k]= \frac{1}{D(q)}w[k] = \frac{1}{1+q^{-1}\overline{D}(q)}w[k]_{\textstyle \raisebox{2pt}{,}}
	\end{align}
\end{subequations}
wherein $D(q)$ represents an FIR filter of order $L_D$.
Similarly, the loudspeaker signal $l[k]$ is a processed version of the microphone signal $m[k]$ through $G(q)$ and $\hat{F}_0(q)$ as 
\begin{equation} \label{eq:l}
	l[k] = -G(q)\hat{F}_0(q)l[k] + G(q)m[k]_{\textstyle \raisebox{2pt}{.}}
\end{equation}
Herein, $G(q)$ represents the forward path processing, and $\hat{F}_0(q)$ represents an FIR filter of order $L_{\hat{F}}$ designed to mimic $F(q)$ in order to remove the acoustic coupling between loudspeaker and microphone. The forward path $G(q)$ additionally contains a delay $\alpha$ of (at least) $1$ sample to avoid an algebraic closed loop, i.e., $G(q)=q^{-\alpha}\overline{G}(q)$. 

\section{TWO-CHANNEL ADAPTIVE FEEDBACK CANCELLER} \label{sec:two_channel_adaptive_feedback_canceller}
The goal of PEM-based identification is to identify $F(q)$ based on observations of $m[k]$ and $l[k]$ \cite{spriet_adaptive_2005,hellgren_bias_2001}, e.g., leading to the 2ch-AFC algorithm of \cite{spriet_adaptive_2005}. This identification results in an FIR estimate $\hat{F}(q)$ of order $L_{\hat{F}}$ for $F(q)$ that can be used for $\hat{F}_0(q)$. To this end, a prediction error $\epsilon_p[k]\in\mathbb{R}$ is defined as \cite{spriet_adaptive_2005}
\begin{subequations}\label{eq:error_q}
	\begin{align}
	\epsilon_p[k] &= B(q)l[k] + A(q)m[k]\\ 
	&= \Vec{b}^\top\underline{\Vec{l}}[k] + m[k] + \overline{\Vec{a}}^\top\underline{\Vec{m}}[k-1]_{\textstyle \raisebox{2pt}{,}}
	\end{align}
\end{subequations}
with 
\begin{subequations}
	\begin{align} 
		B(q)&=b[0]+b[1]q^{-1}+\cdots+b[L_B-1]q^{-L_B+1}\\
		A(q)&=1+a[1]q^{-1}+\cdots+a[L_A-1]q^{-L_A+1}\\
		&= 1 + q^{-1}\overline{A}(q) _{\textstyle \raisebox{2pt}{,}}
	\end{align}
\end{subequations}
and with
\begin{subequations}
	\begin{align}
		\underline{\Vec{m}}[k-1] &= \begin{bmatrix}m[k-1] & \cdots & m[k-L_A+1]\end{bmatrix}^\top\\
		\underline{\Vec{l}}[k] &= \begin{bmatrix}l[k] & \cdots & l[k-L_B+1]\end{bmatrix}^\top_{\textstyle \raisebox{2pt}{.}}
	\end{align}
\end{subequations}
The filter coefficients of $A(q)$ and $B(q)$ are thus contained within $\Vec{a}=\begin{bmatrix}1 & \overline{\Vec{a}}^\top\end{bmatrix}^\top$ and $\Vec{b}$ respectively, and $L_B= L_{\hat{F}}+L_A-1$, $L_A\geq L_D$, and $L_{\hat{F}} \geq L_F$. Additionally, the first coefficient of $A(q)$ is set to one to avoid the trivial solution $A(q)=B(q)=0$ when minimising the prediction error, which does not provide a useful identification\cite{spriet_adaptive_2005}. Under certain identifiability conditions for $G(q)$ (cfr. Section \ref{sec:identifiability}), minimising the mean squared error, i.e.,
\begin{equation} \label{eq:cost}
	\min_{A(q),B(q)} \mathbb{E}\{\epsilon_p[k]^2\}_{\textstyle \raisebox{2pt}{,}}
\end{equation}
yields the unique solution $A(q)=D(q)$ and $B(q)=-A(q)F(q)$.

As $\epsilon_p[k]$ in (\ref{eq:error_q}) is linear in the parameters $\{\Vec{a},\Vec{b}\}$, optimisation of (\ref{eq:cost}) is achieved using linear least squares with the following solution \cite{spriet_adaptive_2005}
\begin{equation} \label{eq:solution_2chAFC}
	\begin{bmatrix}
	\overline{\Vec{a}} \\  \Vec{b}
	\end{bmatrix} = -R^{-1}\Vec{r}_{\textstyle \raisebox{2pt}{,}}
\end{equation}
and with the following correlation matrix and vector
\begin{equation} \label{eq:R}
		R = \begin{bmatrix}R_{\underline{m}\underline{m}} & R_{\underline{m}\underline{l}}\\ R_{\underline{l}\underline{m}} & R_{\underline{l}\underline{l}}\end{bmatrix} \quad \text{and} \quad  
		\Vec{r} = \begin{bmatrix}\Vec{r}_{\underline{m}m}\\\Vec{r}_{\underline{l}m}\end{bmatrix}_{\textstyle \raisebox{2pt}{,}}
\end{equation}
with
\begin{equation}
	\begin{aligned}
		&R_{\underline{m}\underline{m}} = \mathbb{E}\left\{\underline{\Vec{m}}[k-1]\underline{\Vec{m}}[k-1]^\top\right\}_{\textstyle \raisebox{2pt}{,}} R_{\underline{l}\underline{l}} = \mathbb{E}\left\{\underline{\Vec{l}}[k]\underline{\Vec{l}}[k]^\top\right\}_{\textstyle \raisebox{2pt}{,}}\\
		&R_{\underline{l}\underline{m}} = \mathbb{E}\left\{\underline{\Vec{l}}[k]\underline{\Vec{m}}[k-1]^\top\right\}_{\textstyle \raisebox{2pt}{,}}
		R_{\underline{m}\underline{l}} = \mathbb{E}\left\{\underline{\Vec{m}}[k-1]\underline{\Vec{l}}[k]^\top\right\}_{\textstyle \raisebox{2pt}{,}}\\
		&\Vec{r}_{\underline{m}m} = \mathbb{E}\left\{\underline{\Vec{m}}[k-1]m[k]\right\}_{\textstyle \raisebox{2pt}{,}}
		\Vec{r}_{\underline{l}m} = \mathbb{E}\left\{\underline{\Vec{l}}[k]m[k]\right\}_{\textstyle \raisebox{2pt}{.}}
	\end{aligned}
\end{equation}
Finally, since $L_{\hat{F}}>L_F$, $\hat{F}(q)=F(q)$ can be recovered from $B(q)$ by filtering $B(q)$ with $\frac{-1}{A(q)}$ as \cite{spriet_adaptive_2005}
\begin{equation} \label{eq:F_from_B}
\hat{F}(q)=\frac{-1}{A(q)}B(q)=F(q){\textstyle \raisebox{2pt}{.}}
\end{equation} 
The 2ch-AFC algorithm thus solves a least squares problem (\ref{eq:solution_2chAFC}), followed by an inverse filtering operation (\ref{eq:F_from_B}) to estimate the feedback path $\hat{F}(q)$ \cite{spriet_adaptive_2005}. Due to the equivalence of (\ref{eq:F_from_B}), other PEM-based feedback cancellation algorithms can be formulated, e.g., by optimising (\ref{eq:cost}) directly to $\{\!A(q), \hat{F}(q)\!\}$, rather than to $\{\!A(q), B(q)\!\}$ \cite{spriet_adaptive_2005,rombouts_acoustic_2006}. 

\section{IDENTIFIABILITY} \label{sec:identifiability}
In \cite{spriet_adaptive_2005}, it is shown that the PEM-based optimisation of (\ref{eq:cost}) results in $\hat{F}(q)=F(q)$ if $G(q)$ contains a delay $\alpha$ that is equal to or exceeds $L_A$, i.e., $\alpha\geq L_A$. This proof based on (\ref{eq:cost}) applies to all PEM-based feedback cancellation algorithms, thus including the 2ch-AFC algorithm \cite{spriet_adaptive_2005}. This condition will be revisited in Section \ref{subsec:delay_condition}. In this paper, it is then shown that, specifically for the 2ch-AFC algorithm, this identifiability condition can be generalised to $L_{G_{\text{N}}}>L_A$ in Section \ref{subsec:invertibility_condition}.

\subsection{DELAY-BASED CONDITION} \label{subsec:delay_condition}
Using (\ref{eq:m}) and (\ref{eq:l}), the following closed-loop transfer function can be defined to rewrite the loudspeaker signal $l[k]$ as a function of the input signal $s[k]$
\begin{equation} \label{eq:closed_loop_l}
	l[k] = \frac{G(q)}{1-G(q)(F(q)-\hat{F}_0(q))}s[k]_{\textstyle \raisebox{2pt}{,}}
\end{equation}
Assuming $G(q)$ contains a delay $\alpha$, i.e., $G(q)=q^{-\alpha}\overline{G}(q)$ and using (\ref{eq:closed_loop_l}), (\ref{eq:error_q}) can be rewritten as \cite{spriet_adaptive_2005}
\begin{subequations} 
	\begin{align}
	\epsilon_p[k] &= \left(A(q)+q^{-\alpha}\overline{\Gamma}(q)\right)H(q)w[k] \label{eq:error_q_gamma}\\
\overline{\Gamma}(q)&=\frac{\overline{G}(q)(B(q)+A(q)F(q))}{1-G(q)(F(q)-\hat{F}_0(q))}_{\textstyle \raisebox{2pt}{.}}
	\end{align}
\end{subequations} 
Since the first coefficient of $A(q)$ is $1$, as defined in Section \ref{sec:signal_model}, (\ref{eq:cost}) with (\ref{eq:error_q_gamma}) is a linear predictive coding (LPC) of $s[k]$ with solution \cite{spriet_adaptive_2005}
\begin{equation} \label{eq:LPC}
(A(q)+q^{-\alpha}\overline{\Gamma}(q))=H(q)^{-1}=D(q)_{\textstyle \raisebox{2pt}{.}}
\end{equation}
Furthermore, since $\overline{\Gamma}(q)$ can be rewritten as an infinitely long FIR filter $\overline{\Gamma}(q)=\overline{\gamma}[0]+\overline{\gamma}[1]q^{-1}+\cdots$, when $\alpha \geq L_A$, the coefficients of $A(q)$ and $q^{-\alpha}\overline{\Gamma}(q)$ do not overlap, such that (\ref{eq:LPC}) has the unique solution \cite{spriet_adaptive_2005}
\begin{equation}
	A(q)=H(q)^{-1} \ \text{and} \ \overline{\Gamma}(q)=0_{\textstyle \raisebox{2pt}{,}}
\end{equation}
which is only satisfied for \cite{spriet_adaptive_2005}
\begin{equation}
	A(q)=H(q)^{-1} \ \text{and} \ B(q)=-A(q)F(q)_{\textstyle \raisebox{2pt}{.}}
\end{equation}
Consequently, the PEM-based optimisation of (\ref{eq:cost}) results in identification of $F(q)$, i.e., $\hat{F}(q)=F(q)$ when the forward path delay $\alpha$ exceeds $L_A$, i.e.,\footnote{If the feedback path $F(q)$ contains a delay $\beta$, this condition can be modified to $\alpha+\beta\geq L_A$ \cite{spriet_adaptive_2005}. This delay $\beta$ has been omitted here to highlight the equivalence between the results of Section \ref{subsec:delay_condition} and Section \ref{subsec:invertibility_condition}, and was also omitted in the experiments of \cite{spriet_adaptive_2005}.}
\begin{equation}
	\boxed{\alpha \geq L_A}_{\textstyle \raisebox{2pt}{.}} \label{eq:condition_delay}
\end{equation}

\subsection{INVERTIBILITY-BASED CONDITION} \label{subsec:invertibility_condition}
In Section \ref{subsec:invertibility_condition_forward_path} the forward path representation is detailed, using which the invertibility-based condition is derived, specifically for the 2ch-AFC algorithm, in Section \ref{subsec:invertibility_condition_proof}. Finally, in Section \ref{subsec:invertibility_condition_comparison} this invertibility-based condition is compared against other identifiability conditions \cite{boukis_toward_2007,spriet_adaptive_2005,hellgren_bias_2001,benesty_advances_2001}.
\subsubsection{FORWARD PATH REPRESENTATION} \label{subsec:invertibility_condition_forward_path}
In \cite{spriet_adaptive_2005}, $G(q)$ was assumed to be a general linear filter. Its structure was not explicitly considered due to the focus on the forward path delay $\alpha$. However, without loss of generality, the forward path $G(q)$ can be rewritten as an IIR filter of the following structure
\begin{subequations}
	\begin{align}
	G(q) &= \frac{G_\text{N}(q)}{G_\text{D}(q)} = \frac{q^{-1}\overline{G}_\text{N}(q)}{1+q^{-1}\overline{G}_\text{D}(q)} \label{eq:def_G}\\
	&=\frac{g_{\text{N}}[1]q^{-1}+\cdots+g_{\text{N}}[L_{G_{\text{N}}}-1]q^{-L_{G_{\text{N}}}+1}}{1+g_{\text{D}}[1]q^{-1}+\cdots+g_{\text{D}}[L_{G_{\text{D}}}-1]q^{-L_{G_{\text{D}}}+1}}_{\textstyle \raisebox{2pt}{,}}
	\end{align}
\end{subequations}
wherein the forward path delay of one sample is made explicit, and wherein $G_{\text{N}}(q)$ of order $L_{G_{\text{N}}}$ represents an FIR feedforward filter and $G_{\text{D}}(q)$ of order $L_{G_{\text{D}}}$ represents an FIR feedback filter. The additional delay of $\alpha-1$ samples is assumed to be captured by the coefficients of $\overline{G}_\text{N}(q)$. 

The filter coefficients of ${G}_\text{N}(q)$ and ${G}_\text{D}(q)$ (or $\overline{G}_\text{N}(q)$ and $\overline{G}_\text{D}(q)$) can alternatively be collected into $\Vec{g}_\text{N}\in\mathbb{R}^{L_{G_{\text{N}}}\times 1}$ and $\Vec{g}_\text{D}\in\mathbb{R}^{L_{G_{\text{D}}}\times 1}$ (or $\overline{\Vec{g}}_\text{N}\in\mathbb{R}^{L_{G_{\text{N}}}-1\times 1}$ and $\overline{\Vec{g}}_\text{D}\in\mathbb{R}^{L_{G_{\text{D}}}-1\times 1}$), respectively, as
\begin{subequations}
	\begin{align}
		\Vec{g}_\text{N} &= \begin{bmatrix}0 & g_{\text{N}}[1] & \cdots & g_{\text{N}}[L_{G_\text{N}}-1]\end{bmatrix}^\top = \begin{bmatrix}0 & \overline{\Vec{g}}_{\text{N}}^\top\end{bmatrix}^\top\\
		\Vec{g}_\text{D} &= \begin{bmatrix}1 & g_{\text{D}}[1] & \cdots & g_{\text{D}}[L_{G_\text{D}}-1]\end{bmatrix}^\top = \begin{bmatrix}1 & \overline{\Vec{g}}_{\text{D}}^\top\end{bmatrix}^\top_{\textstyle \raisebox{2pt}{.}}
	\end{align}
\end{subequations} 

Finally, $G_{\text{N}}(q)$ and $G_{\text{D}}(q)$ are assumed to share no common roots, such that the forward path $G(q)$ in (\ref{eq:def_G}) corresponds to a minimal realisation for which no pole-zero cancellation occurs.
\subsubsection{ANALYTICAL FORMULATION} \label{subsec:invertibility_condition_proof}
In a first step it will be shown that a necessary and sufficient condition for identifiability corresponds to $R$ in (\ref{eq:R}) being of full rank. In a second step, it will be shown that this condition holds true when the order of the forward path feedforward filter
exceeds the order of the AR model, thereby providing a sufficient condition for identifiability. It is reiterated that this proof applies only to the 2ch-AFC algorithm.

Equation (\ref{eq:m}), containing the microphone signal model, can be rewritten by multiplying both sides with $D(q)=1+q^{-1}\overline{D}(q)$, resulting in
\begin{subequations}
	\begin{align} 
		m[k] &= F(q)l[k] + \frac{1}{1+q^{-1}\overline{D}(q)}w[k]\\
		&= D(q)F(q)l[k] + w[k] - \overline{D}(q)m[k-1]_{\textstyle \raisebox{2pt}{.}}\label{eq:m_rewritten1}
	\end{align}
\end{subequations}		
Replacing the filter operations in (\ref{eq:m_rewritten1}) with a vector-based notation then results in  
\begin{equation}	
	m[k] = \left(\Vec{d}*\Vec{f}\right)_{\text{ext}}^\top\underline{\Vec{l}}[k] + w[k] - \overline{\Vec{d}}_{\text{ext}}^\top\underline{\Vec{m}}[k-1]_{\textstyle \raisebox{2pt}{.}} \label{eq:m_rewritten}
\end{equation}
Herein, $*$ represents the convolution operation, and the subscript $._{\text{ext}}$ denotes zero-padding as
\begin{subequations}
	\begin{align} 
	\left(\Vec{d}*\Vec{f}\right)_{\text{ext}} &= \begin{bmatrix}\left(\Vec{d}*\Vec{f}\right)^\top & \Vec{0}^\top_{L_B-(L_F+L_D-1)\times 1} \end{bmatrix}^\top\\
	\overline{\Vec{d}}_{\text{ext}} &= \begin{bmatrix}\overline{\Vec{d}}^\top & \Vec{0}^\top_{L_A-L_D\times 1}\end{bmatrix}^\top_{\textstyle \raisebox{2pt}{,}}
	\end{align}
\end{subequations}
which is added since the algorithm parameters $L_A$ and $L_B$ can be chosen larger than the true $L_D$ and $L_A+L_F-1$ as detailed in Section \ref{sec:signal_model}.

Since $G(q)$ contains a delay $\alpha \geq 1$ and $w[k]$ is white noise, $\mathbb{E}\{l[k]w[k]\}=0$ and $\mathbb{E}\{m[k-1]w[k]\}=0$, such that $\Vec{r}_{\underline{m}m}$ and $\Vec{r}_{\underline{l}m}$ can be rewritten solely using (\ref{eq:m_rewritten}) as 
\begin{subequations} \label{eq:r_rewritten}
	\begin{align} 
		\Vec{r}_{\underline{m}m} &= R_{\underline{m}\underline{l}}\left(\Vec{d}*\Vec{f}\right)_{\text{ext}} - R_{\underline{m}\underline{m}}\overline{\Vec{d}}_{\text{ext}}\\
		\Vec{r}_{\underline{l}m} &= R_{\underline{l}\underline{l}}\left(\Vec{d}*\Vec{f}\right)_{\text{ext}} - R_{\underline{l}\underline{m}}\overline{\Vec{d}}_{{\text{ext}}\textstyle \raisebox{2pt}{.}}
	\end{align}
\end{subequations}

Plugging (\ref{eq:r_rewritten}) into (\ref{eq:solution_2chAFC}) yields
\begin{subequations}
	\begin{align} 
	\begin{bmatrix}
	\overline{\Vec{a}}\\ \Vec{b}
	\end{bmatrix} &= \begin{bmatrix}R_{\underline{m}\underline{m}} & R_{\underline{m}\underline{l}}\\ R_{\underline{l}\underline{m}} & R_{\underline{l}\underline{l}}\end{bmatrix}^{-1}\begin{bmatrix}R_{\underline{m}\underline{m}} & R_{\underline{m}\underline{l}}\\ R_{\underline{l}\underline{m}} & R_{\underline{l}\underline{l}}\end{bmatrix}\begin{bmatrix}\overline{\Vec{d}}_{\text{ext}}\\ - \left(\Vec{d}*\Vec{f}\right)_{\text{ext}}\end{bmatrix}\label{eq:invertibility_proof}\\
	&= \begin{bmatrix}\overline{\Vec{d}}_{\text{ext}}\\ - \left(\Vec{d}*\Vec{f}\right)_{\text{ext}}\end{bmatrix}_{\textstyle \raisebox{2pt}{,}}	
	\end{align}
\end{subequations}
such that identifiability, i.e., $A(q)=D(q)$ and $B(q)=-A(q)F(q)$, is achieved when $R$ is invertible. Alternatively formulated, a necessary and sufficient condition for identifiability is as follows
\begin{equation}
\boxed{\begin{array}{l}\text{The 2ch-AFC combined input vector}\\\underline{\Vec{i}}[k]=\begin{bmatrix}\underline{\Vec{m}}[k-1]^\top & \underline{\Vec{l}}[k]^\top\end{bmatrix}^\top\\ \text{contains linearly independent signals.} \end{array}}_{\textstyle \raisebox{2pt}{,}} \label{eq:extended_vector_def}
\end{equation}

In a second step, to study this property, the loudspeaker signals should be rewritten only as a linear combination of past microphone and loudspeaker signals. To this end, (\ref{eq:l}), containing the loudspeaker signal model, can be rewritten as a function of FIR filters applied to $l[k]$, and $m[k]$ by multiplying both sides with $G_{\text{D}}(q)=1+q^{-1}\overline{G}_{\text{D}}(q)$
\begin{subequations}
	\begin{align} 
		&l[k] = -G(q)\hat{F}_0(q)l[k] + G(q)m[k]\\
		&= \frac{-q^{-1}\overline{G}_{\text{N}}(q)}{1+q^{-1}\overline{G}_{\text{D}}(q)}\hat{F}_0(q)l[k] + \frac{q^{-1}\overline{G}_{\text{N}}(q)}{1+q^{-1}\overline{G}_{\text{D}}(q)}m[k]\\
		&= -\left(\overline{G}_{\text{N}}(q)\hat{F}_0(q)+\overline{G}_{\text{D}}(q)\right)l[k-1] + G_{\text{N}}(q)m[k]_{\textstyle \raisebox{2pt}{,}} \label{eq:l_rewritten1}
	\end{align}
\end{subequations}
Replacing the filter operations in (\ref{eq:l_rewritten1}) with a vector-based notation then results in  
\begin{equation}
	\begin{aligned}
		&l[k]= \\ 
		& -\left((\overline{\Vec{g}}_{\text{N}} * \hat{\Vec{f}}_0) + \overline{\Vec{g}}_{\text{D}}\right)^\top\begin{bmatrix}l[k-1]\\\vdots\\l[k-\max(L_{G_{\text{N}}}+L_{\hat{F}}-1,L_{G_{\text{D}}}) + 1]\end{bmatrix}^\top  \\
		& \quad + \overline{\Vec{g}}_{\text{N}}^\top \begin{bmatrix}m[k-1]\\ \vdots \\ m[k-L_{G_{\text{N}}}+1]\end{bmatrix}^\top_{\textstyle \raisebox{2pt}{.}} \label{eq:l_rewritten}
	\end{aligned}
\end{equation}

When $L_{G_{\text{N}}} > L_A$, the loudspeaker signals in $\underline{\Vec{i}}[k]$ are a function of loudspeaker and microphone signals not readily available to the algorithm as these are not present in the input vector of the 2ch-AFC algorithm $\underline{\Vec{i}}[k]$. Indeed, $\underline{\Vec{l}}[k]$ then contains $L_B=L_{\hat{F}}+L_A-1 < L_{\hat{F}}+L_{G_\text{N}}-1$ samples and $\underline{\Vec{m}}[k-1]$ contains $L_A < L_{G_\text{N}}$ samples. Similarly, when $L_{G_{\text{D}}} - L_{\hat{F}} + 1> L_A$, the loudspeaker signals in $\underline{\Vec{i}}[k]$ are a function of loudspeaker signals not readily available to the algorithm as these are not present in $\underline{\Vec{i}}[k]$ as $\underline{\Vec{l}}[k]$ then contains $L_B=L_{\hat{F}}+L_A-1 < L_{G_{\text{D}}}$ samples. For example, $l[k]$ also depends on $l[k-L_B+2],\cdots,l[k-L_{G_D}]$, unavailable in $\underline{\Vec{i}}[k]$ when $L_B < L_{G_{\text{D}}}$. Additionally, with (\ref{eq:m_rewritten}), the system is excited by white noise $w[k]$ such that microphone and loudspeaker signals are filtered versions thereof, and the system is persistently excited. Thus, none of the loudspeaker and microphone signals in $\underline{\Vec{i}}[k]$ can be rewritten as a linear combination of the other elements in $\underline{\Vec{i}}[k]$ when either $L_{G_{\text{N}}} > L_A$ or $L_{G_{\text{D}}} - L_{\hat{F}} + 1> L_A$.

In conclusion, the PEM-based optimisation using the 2ch-AFC algorithm of (\ref{eq:solution_2chAFC}) results in identification of the feedback path $F(q)$, i.e., $\hat{F}(q)=F(q)$ when
\begin{equation}
	\boxed{\max(L_{G_{\text{N}}},L_{G_{\text{D}}} - L_{\hat{F}} + 1) > L_A}_{\textstyle \raisebox{2pt}{.}} \label{eq:condition_length1}
\end{equation}

While the condition $L_{G_{\text{D}}} - L_{\hat{F}} + 1 > L_A$ provides identifiability, when implementing IIR filters in practice typically shorter filter orders are preferred in the denominator, i.e., in practice, $L_{G_{\text{D}}}$ is on the order of ten \cite{boukis_toward_2007}. Furthermore, in hearing instruments $L_{\hat{F}}$ is typically chosen between $64$ and $100$ at a sampling rate $f_s=\SI{16}{k\hertz}$ \cite{sankowsky-rothe_acoustic_2020,spriet_adaptive_2005,lydaki_deep_feedback_2025}, and in PA systems $L_{\hat{F}}$ could increase to $8000$ at a sampling rate $f_s=\SI{16}{k\hertz}$ \cite{rombouts_acoustic_2006}. In contrast, $L_{G_{\text{N}}}$ can typically be of larger order. Consequently, from a practical perspective, $L_{G_{\text{N}}}$ is easier to manipulate, leading to the following sufficient (but not necessary) condition for identifiability
\begin{equation}
	\boxed{L_{G_{\text{N}}} > L_A}_{\textstyle \raisebox{2pt}{,}} \label{eq:condition_length}
\end{equation}
i.e., identifiability is obtained when the order of the forward path feedforward filter exceeds $L_A$.

Nevertheless, in practice, this condition (\ref{eq:condition_length}) could still result in an ill-conditioned identification if $L_{G_{\text{N}}}$ only marginally exceeds $L_A$ and/or if the first $L_{A}$ coefficients of $G_{\text{N}}(q)$ are significantly larger than the other coefficients of $G_{\text{N}}(q)$. To monitor this, the condition number of inversion of $R$ in (\ref{eq:solution_2chAFC}), i.e., $\kappa(R)$, could be used with
\begin{equation} \label{eq:cond_R}
	\kappa(R)=\frac{\sigma_{\text{max}}(R)}{\sigma_{\text{min}}(R)}_{\textstyle \raisebox{2pt}{.}} 
\end{equation}
Herein, $\sigma_{\text{max}}(R)$ and $\sigma_{\text{min}}(R)$ denote the largest and smallest singular value of $R$, respectively \cite{trefethen_numerical_1997}. A larger $\kappa(R)$ indicates a more ill-conditioned identification. 
\subsubsection{RELATION TO EXISTING CONDITIONS} \label{subsec:invertibility_condition_comparison}
The necessary and sufficient condition (\ref{eq:extended_vector_def}) requires linearly independent signals in $\underline{\Vec{i}}[k]$ to achieve identifiability. It can be observed that this necessary and sufficient condition is also satisfied when $G(q)$ is sufficiently time-varying or non-linear as consistent with \cite{hellgren_bias_2001,spriet_adaptive_2005,boukis_toward_2007}, or when the forward path delay is equal to or exceeds the order of the AR model as consistent with \cite{spriet_adaptive_2005}.

The invertibility-based condition (\ref{eq:condition_length}), although only proven for the 2ch-AFC algorithm, additionally provides a more general identifiability condition than the delay-based condition (\ref{eq:condition_delay}). Indeed, when the first $L_A$ coefficients of $G_{\text{N}}(q)$ are chosen to be zero, i.e., 
\begin{equation}
	\begin{bmatrix}g_{\text{N}}[0] & \cdots & g_{\text{N}}[L_A-1]\end{bmatrix}^\top=\Vec{0}_{L_A\times 1{\textstyle \raisebox{2pt}{,}}}
\end{equation}
the delay-based condition is retrieved as a special instance of the invertibility-based condition. In practice, as $L_A$ is typically on the order of ten, rather than requiring that the first several tens of filter coefficients are set to $0$, more flexibility is offered as long as the filter order exceeds $L_A$. As the forward paths in hearing instruments and PA systems that contain the signal processing algorithms typically have several tens or hundreds of filter coefficients \cite{kates_digital_2008,van_waterschoot_fifty_2011}, this condition will generally be satisfied.
 
In \cite{boukis_toward_2007}, identifiability was retrieved when using a sufficiently time-varying all-pass IIR filter as forward path. Nevertheless, given (\ref{eq:condition_length}), in addition to using time-varying all-pass IIR filters \cite{boukis_toward_2007}, stationary all-pass IIR filters can also obtain identifiability, and the time-variety is a sufficient but non-necessary requirement when (\ref{eq:condition_length1}) is satisfied.

Condition (\ref{eq:condition_length}) is furthermore similar to the frequency-domain condition postulated in \cite{hellgren_bias_2001}, where PEM-based identifiability is shown to be achieved when the $2\times 2$ matrix of loudspeaker and white noise (cross-)power spectral densities, formulated in the continuous-time Fourier transform domain, is of full rank. In \cite{hellgren_bias_2001}, it is consequently argued that identifiability is obtained for non-linear or time-varying forward path processing $G(q)$. In contrast, by using the 2ch-AFC algorithm as an explicit starting point, and leveraging the finite length of the feedback path $F(q)$ and AR model denominator $D(q)$, the more general condition (\ref{eq:condition_length}) is obtained, be it specifically for the 2ch-AFC algorithm. Additionally, due to the formulation in the continuous-time Fourier transform domain, practical evaluation of this condition is challenging, as opposed to evaluating the condition number of time-domain correlation matrices as in (\ref{eq:cond_R}).

The conditioning of the identification as described in Section \ref{subsec:invertibility_condition_forward_path} is similar to the conditioning for the multi-channel acoustic echo cancellation (AEC) problem when using undermodelled AEC filters \cite{benesty_better_1998}. Indeed, in such AEC problems, the loudspeaker correlation matrix, containing the loudspeaker signals from all loudspeakers, is to be inverted similar to $R$ in (\ref{eq:solution_2chAFC}). When the AEC filters are long enough to model the far-end room transfer functions from the source to the microphones, the loudspeaker signals in this loudspeaker correlation matrix are linearly dependent, and the matrix is not invertible. Yet, when the AEC filters are too short to model these far-end room transfer functions, the loudspeaker correlation matrix becomes invertible, but can still be ill-conditioned \cite{benesty_better_1998}.

\section{EXPERIMENT DESIGN} \label{sec:experiment_design}
The identifiability conditions of Section \ref{sec:identifiability} are experimentally validated. To this end, the acoustic scenarios are described in Section \ref{subsec:acoustic_scenario}, the algorithm settings in Section \ref{subsec:algorithm_settings}, and the performance measures in Section \ref{subsec:performance_measures}.

\subsection{ACOUSTIC SCENARIOS} \label{subsec:acoustic_scenario}
Three feedback paths of $64$ samples ($f_s = \SI{16}{k\hertz}$) are examined, as measured using a behind-the-ear (BTE) hearing instrument\cite{spriet_evaluation_2010}, i.e., a feedback path with a hand close to the BTE, a hand far from the BTE, and paper close to the BTE. 

Either stationary speech-shaped noise obtained by passing white Gaussian noise through an AR model with $L_D=10$ or $L_D=20$, or real-life speech consisting of hearing in noise test (HINT) sentences \cite{nilsson_development_1994} are used with a length of $\SI{45}{\second}$ and $f_s = \SI{16}{k\hertz}$. The AR models have been created through linear predictive coding of the HINT sentences, and the overall power of the speech-shaped noise is of the same power as the speech of the HINT sentences. In Section \ref{subsec:noise_influence}, babble noise is added to the speech input to study the effect of the signal-to-noise ratio (SNR) on the identification \cite{auditecAuditoryTestsRevised1997}.

The forward path gain is set such that the overall system gain is $\SI{3}{\decibel}$ below the maximum stable gain (MSG) \cite{van_waterschoot_fifty_2011}
\begin{equation}
	\text{MSG [\SI{}{\decibel}]} = -20\text{log}_{10}\!\left(\max_{\omega \in \mathcal{P}}|G(\omega)F(\omega)|\right)_{\textstyle \raisebox{2pt}{,}} 
\end{equation}
with radial frequency $\omega$ and $\mathcal{P}\! =\! \left\{\omega|\angle G(\omega)F(\omega)\!=\!n2\pi\right\}$ the radial frequencies for which the system can become unstable. 

\subsection{ALGORITHM SETTINGS} \label{subsec:algorithm_settings}
The following forward path filters are considered:
\begin{itemize}
\item $G_{\text{FIR}}(q)$: An FIR filter of $L_{G_{\text{N}}}$ samples, with a leading delay of $\alpha\geq 1$ samples, i.e.,
\begin{equation}
\Vec{g}_{\text{FIR}} \!=\! g\begin{bmatrix}
\multicolumn{3}{c}{\smash[b]{\underbrace{0 \!\quad\! \cdots \!\quad\! 0}_{\alpha}}} \!&\! \smash[b]{\underbrace{x_0 \!\quad\! \cdots \!\quad\! x_{L_{G_{\text{N}}}-\alpha-1}}_{L_{G_{\text{N}}}-\alpha}}
\end{bmatrix}^\top\in \mathbb{R}^{L_{G_{\text{N}}}\times 1}_{\textstyle \raisebox{2pt}{,}}
\end{equation}\\
where the $x_j$'s ($j\in[0,L_{G_{\text{N}}}-\alpha-1]$) correspond to random coefficients drawn from a standard normal distribution. Additionally, $g$ is chosen to achieve the gain as defined in Section \ref{subsec:acoustic_scenario}.
\item $G_{\text{IIR-AP}}(q)$: An all-pass IIR filter with $L_{G_{\text{N}}}$ samples for the filter $G_{\text{N},\text{IIR-AP}}(q)$ in the numerator of $G_{\text{IIR-AP}}(q)$, with a leading delay of $\alpha\geq1$ samples, i.e.,
\begin{equation}
\Vec{g}_{\text{N},\text{IIR-AP}} \!=\! g\begin{bmatrix}
\multicolumn{3}{c}{\smash[b]{\underbrace{0 \!\quad\! \cdots \!\quad\! 0}_{\alpha}}} \!&\! \smash[b]{\underbrace{x_0 \!\quad\! \cdots \!\quad\! x_{L_{G_{\text{N}}-\alpha-2}}}_{L_{G_{\text{N}}}-\alpha-1}} \!&\! 1
\end{bmatrix}^\top\in \mathbb{R}^{L_{G_{\text{N}}}\times 1}_{\textstyle \raisebox{2pt}{,}}
\end{equation} \\
and a denominator filter $G_{\text{D},\text{IIR-AP}}(q)$ of $L_{G_{\text{D}}}$ samples 
\begin{subequations}
\begin{align}
\Vec{g}_{\text{D},\text{IIR-AP}} &= P\Vec{g}_{\text{N},\text{IIR-AP}} \quad \ \ \in \mathbb{R}^{L_{G_{\text{D}}}\times 1}\\
P &= \begin{bmatrix}0 & \cdots & 1\\ \vdots & \udots & \vdots\\ 1 & \cdots & 0  \end{bmatrix} \in \mathbb{R}^{L_{G_{\text{N}}}\times L_{G_{\text{N}}}}_{\textstyle \raisebox{2pt}{.}}
\end{align}
\end{subequations}	
\item $G_{\text{Delay1}}(q)$: An FIR filter corresponding to a pure delay and gain where the delay equals $L_A$, i.e., a forward path of the following type
\begin{equation}
\Vec{g}_{\text{Delay1}} = g\begin{bmatrix}
\multicolumn{3}{c}{\smash[b]{\underbrace{0 \quad \cdots \quad 0}_{L_A}}} & 1
\end{bmatrix}^\top\in \mathbb{R}^{L_A+1\times 1}_{\textstyle \raisebox{2pt}{.}}
\end{equation}
\ \\This FIR all-pass filter is the shortest FIR filter satisfying the delay constraint of Section \ref{subsec:delay_condition}, is commonly used for AFC algorithms \cite{van_waterschoot_fifty_2011,lydaki_deep_feedback_2025, spriet_adaptive_2005,rombouts_acoustic_2006}, and serves as a first baseline identifiability condition.
\item $G_{\text{Delay2}}(q)$: An FIR filter with a forward path corresponding to a pure delay and gain of $L_{G_{\text{N}}}$ samples, i.e., an FIR filter of the same length as $G_{\text{FIR}}(q)$ but with $L_{G_{\text{N}}}-1$ zero-coefficients and a pure gain as the final coefficient. It serves as a second baseline identifiability condition. Thus, $\Vec{g}_{\text{Delay2}}$ corresponds to
\begin{equation}
\Vec{g}_{\text{Delay2}}= g\begin{bmatrix}
\multicolumn{3}{c}{\smash[b]{\underbrace{0 \quad \cdots \quad 0}_{L_{G_{\text{N}}}-1}}} & 1
\end{bmatrix}^\top\in \mathbb{R}^{L_{G_{\text{N}}}\times 1}_{\textstyle \raisebox{2pt}{.}}
\end{equation}
\end{itemize}
\begin{figure*}[h!tbp]
    \centering
    \subfigure[ASG when using $G_{\text{FIR}}$(q), $G_{\text{Delay1}}(q)$, and $G_{\text{Delay2}}(q)$.]{
        \includegraphics[width=0.785\linewidth]{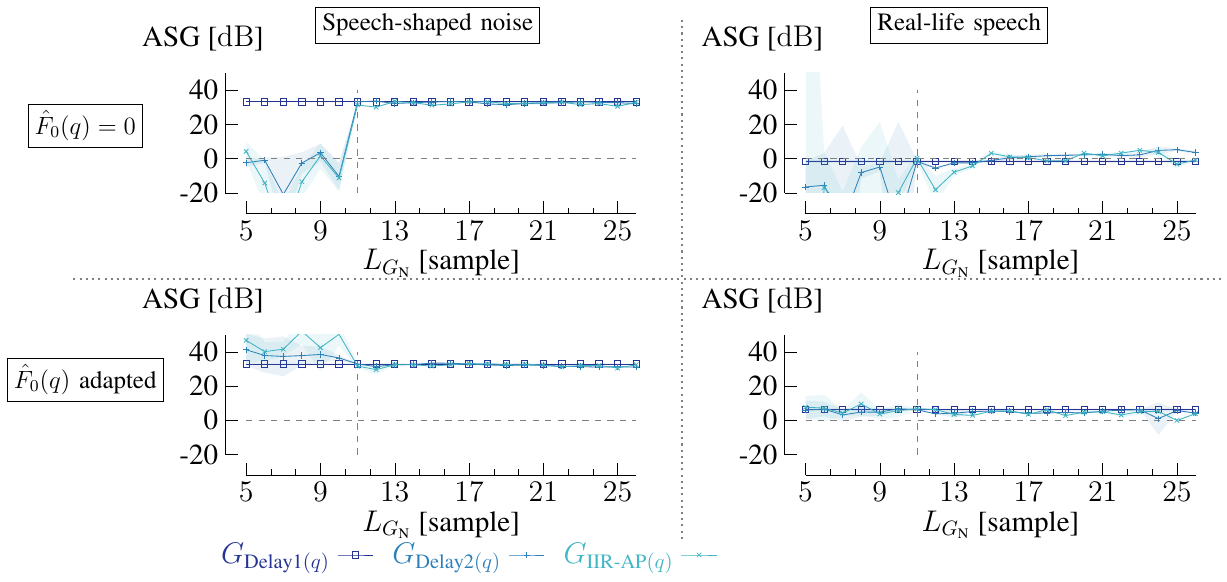}
        \label{fig:IIR_comp}}
    \subfigure[Condition number $\kappa(R)$ when using $G_{\text{FIR}}$(q), $G_{\text{Delay1}}(q)$, and $G_{\text{Delay2}}(q)$.]{
        \includegraphics[width=0.785\linewidth]{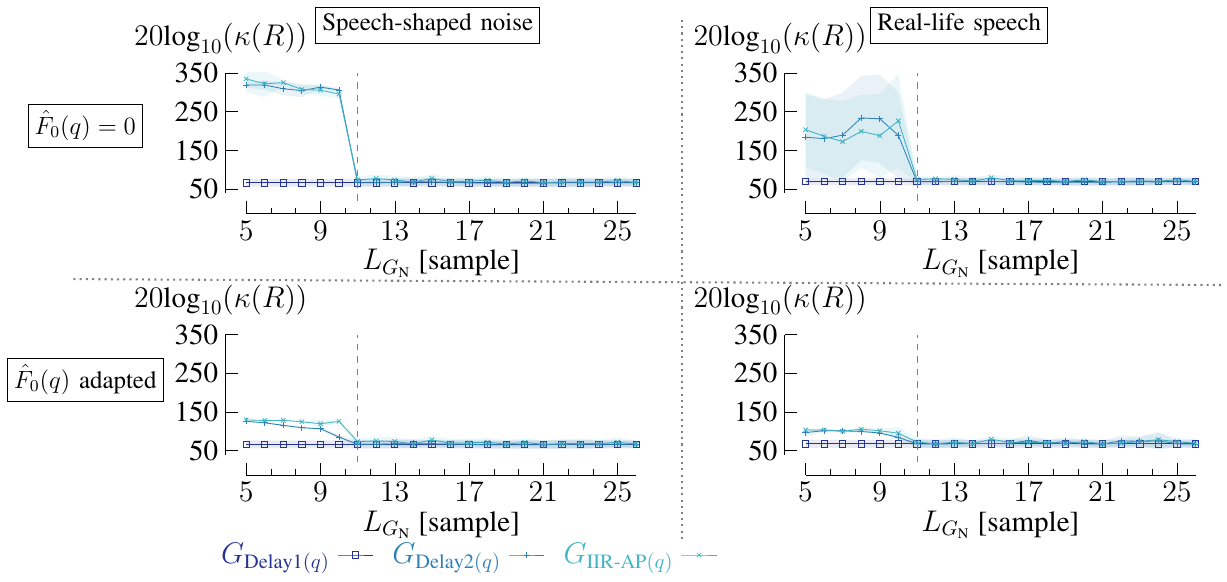}
        \label{fig:IIR_cond}}		
	\caption{Identifiability for $G_{\text{IIR-AP}}(q)$ with $\alpha=1$ in terms of added stable gain (ASG) (Fig. \ref{fig:IIR_comp}) and condition number $\kappa(R)$ (Fig. \ref{fig:IIR_cond}). Corresponding to the theoretical condition, identifiability is obtained when $L_{G_{\text{N}}}>L_A$, as indicated by the dashed line. The solid line and shading indicate the mean and the standard deviation across the feedback paths, respectively.}
	\label{fig:IIR}
\end{figure*}
\vspace{1em}
Two scenarios are explored. In a first scenario, no feedback canceller $\hat{F}_0(q)$ is inserted to achieve offline identification of the feedback path $F(q)$, i.e., the identified feedback path is not inserted in the closed loop, thus $\hat{F}_0(q)=0$. The correlation matrix $R$ and vector $\Vec{r}$ of (\ref{eq:R}) are computed using sample averages. In a second scenario, $\hat{F}_0(q)$ is replaced by the current estimate $\hat{F}(q)$ at each discrete-time instant, to achieve a recursive cancellation and to aid the identifiability of $F(q)$ through time-variability \cite{spriet_adaptive_2005}. To this end, at each discrete-time instant, the recursive least squares (RLS) algorithm with forgetting factor of $1$ is used to compute (\ref{eq:solution_2chAFC}) adaptively as per \cite{spriet_adaptive_2005}, and the resulting $\hat{F}(q)$ replaces $\hat{F}_0(q)$. To remove any DC offset in the identified feedback path, the mean of the estimated feedback path is subtracted from this estimate, as the DC offset is similarly not present in $F(q)$.

The number of coefficients of $\Vec{a}$, i.e., $L_A$, is either set to $10$ or $20$ depending on the order $L_{\text{D}}$ used, and the number of coefficients of $\Vec{b}$, i.e., $L_B$, is set to $L_A+L_{\hat{F}}-1$ with $L_{\hat{F}}=L_F=64$. To avoid unbounded diverging of the feedback path estimates $\hat{F}(q)$, excessively large estimated feedback path coefficients ($>\!10$) and loudspeaker signal amplitudes ($>\!1$) are clipped. Additionally, when recursively updating the feedback paths, $\hat{F}_0(q)$ is only replaced with $\hat{F}(q)$ after $\SI{1}{\second}$ to allow for initial convergence of $\hat{F}(q)$.

\subsection{PERFORMANCE MEASURES} \label{subsec:performance_measures}
The added stable gain (ASG) defines the added margin of stability when inserting $\hat{F}(q)$ as $\hat{F}_0(q)$, i.e., \cite{van_waterschoot_fifty_2011}
\begin{equation}
	\text{ASG\!\! [\SI{}{\decibel}]} \! = \!\text{MSG}\! - \! 20\text{log}_{10}\!\left(\!\max_{\omega \in \mathcal{P}}\!|G(\omega)(\!F(\omega)\!-\!\hat{F}(\omega)\!)|\!\right)_{\textstyle \raisebox{2pt}{,}}  	
\end{equation}
such that ASG margins larger than zero indicate an increase in the system's stability. Additionally, the conditioning of the identifiability is measured through $\kappa(R)$ of (\ref{eq:cond_R}).

\section{RESULTS AND DISCUSSION} \label{sec:results_and_discussion}
In Section \ref{subsec:iir}, $G_{\text{IIR-AP}}(q)$ is analysed and compared to $G_{\text{Delay1}}(q)$ and $G_{\text{Delay2}}(q)$. In Section \ref{subsec:fir}, a similar comparison is conducted for $G_{\text{FIR}}(q)$. In Section \ref{subsec:noise_influence}, the effect of the SNR of the input signal on the identification is analysed.

\subsection{INFINITE IMPULSE RESPONSE ALL-PASS (IIR-AP)} \label{subsec:iir}
Fig. \ref{fig:IIR} shows the comparison in ASG (Fig. \ref{fig:IIR_comp}) and in $\kappa(R)$ (Fig. \ref{fig:IIR_cond}) when using forward paths $G_{\text{IIR-AP}}(q)$, $G_{\text{Delay1}}(q)$, and $G_{\text{Delay2}}(q)$ for speech-shaped noise with $L_D=10$ and for real-life speech. The forward path $G_{\text{IIR-AP}}(q)$ contains a delay $\alpha=1$, necessary to avoid an algebraic closed-loop. The solid line indicates the mean across the three feedback paths, and the shading indicates the standard deviation.

In line with condition (\ref{eq:condition_length}), identifiability is obtained when $L_{G_{\text{N}}}>L_A=10$ as also indicated by the dashed line in Fig. \ref{fig:IIR}. This identifiability is visible for $G_{\text{IIR-AP}}(q)$ and $G_{\text{Delay2}}(q)$ both from a converged and positive ASG in Fig. \ref{fig:IIR_comp}, and from a lower and converged $\kappa(R)$ when $L_{G_{\text{N}}}>L_A=10$ in Fig. \ref{fig:IIR_cond}. Consequently, $\kappa(R)$ and the ASG do indeed seem to be inversely related to one another. It can also be observed that when adapting $\hat{F}_0(q)$ recursively, the difference in ASG and $\kappa(R)$ for the region $L_{G_{\text{N}}}\leq L_A$ and $L_{G_{\text{N}}} > L_A$ is smaller than when $\hat{F}_0(q)=0$. This is in correspondence with \cite{spriet_adaptive_2005,hellgren_bias_2001,boukis_toward_2007} where time-varying forward paths were shown to aid the identifiability. Finally, in line with the conclusions in \cite{spriet_adaptive_2005}, the ASG is lower in real-life speech scenarios compared to speech-shaped noise scenarios as real-life speech cannot be exactly modelled using an AR model.

\subsection{FINITE IMPULSE RESPONSE (FIR)} \label{subsec:fir}
\begin{figure*}[htbp]
    \centering
    \subfigure[ASG when using $G_{\text{FIR}}$(q) for different forward path delays $\alpha$.]{
        \includegraphics[width=0.785\linewidth]{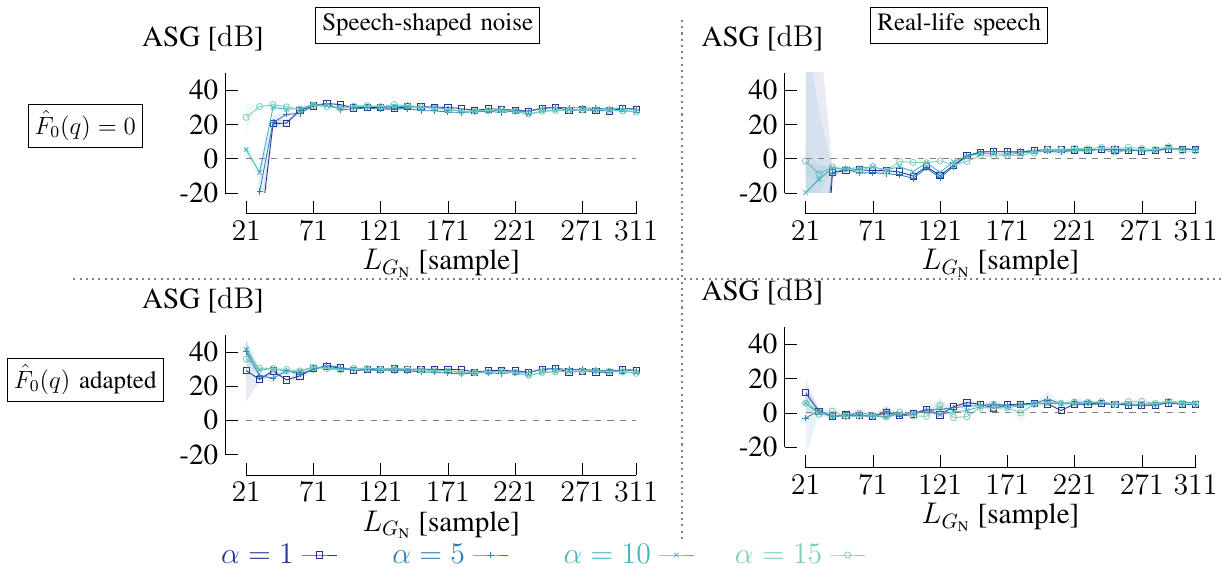}
        \label{fig:FIR_ASG}}	
    \subfigure[Condition number $\kappa(R)$ when using $G_{\text{FIR}}$(q) for different forward path delays $\alpha$.]{
        \includegraphics[width=0.785\linewidth]{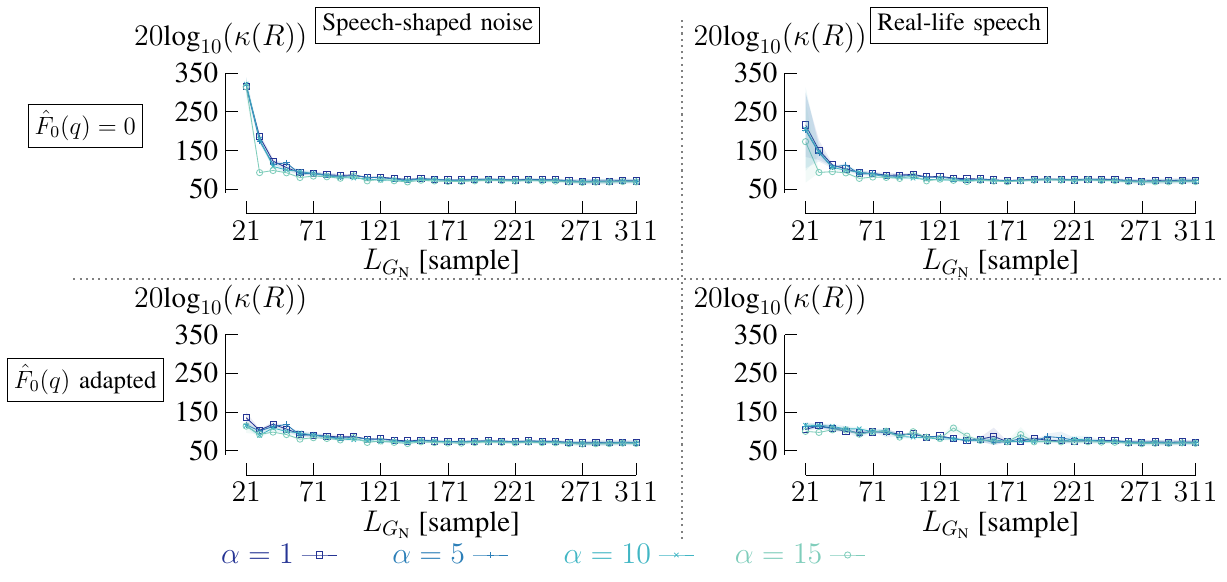}
        \label{fig:FIR_cond}}				
    \subfigure[ASG when using $G_{\text{FIR}}$(q), $G_{\text{Delay1}}(q)$, and $G_{\text{Delay2}}(q)$.]{
        \includegraphics[width=0.785\linewidth]{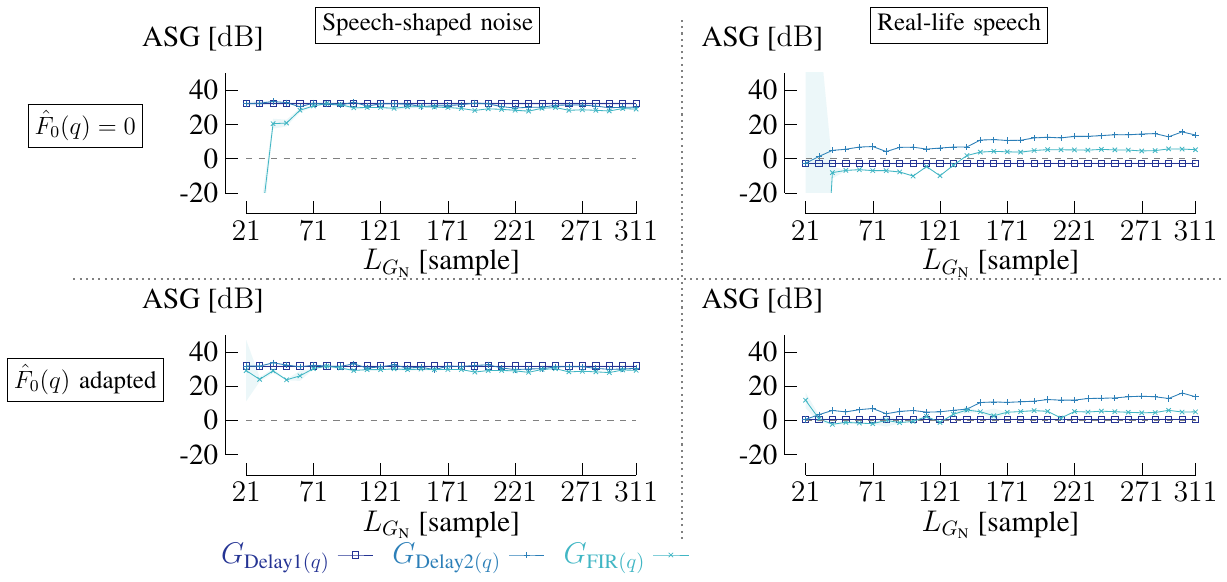}
        \label{fig:FIR_comp}}							
	\caption{Identifiability for $G_{\text{FIR}}(q)$ for different forward path delays $\alpha$ in terms of added stable gain (ASG) (Fig. \ref{fig:FIR_ASG}) and condition number $\kappa(R)$ (Fig. \ref{fig:FIR_cond}), and comparison to $G_{\text{Delay1}}(q)$ and $G_{\text{Delay2}}(q)$ with  $\alpha=1$. In practice, a filter length $L_{G_{\text{N}}}$ larger than the theoretical bound $L_A$ could be required, as measurable through $\kappa(R)$. The solid line and shading indicate the mean and the standard deviation across the feedback paths, respectively.}
	\label{fig:FIR}
\end{figure*}

Fig. \ref{fig:FIR_ASG} shows the ASG when using forward path $G_{\text{FIR}}(q)$ for speech-shaped noise with $L_D=20$ and for real-life speech for different choices of delay $\alpha$ While according to the theory of Section \ref{subsec:invertibility_condition_proof} identifiability should be obtained when $L_{G_{\text{N}}}>L_A=20$ (\ref{eq:condition_length}), on the contrary, the feedback paths are not directly convergent for each curve in Fig. \ref{fig:FIR_ASG}. Indeed, as detailed in Section \ref{subsec:invertibility_condition}, in practice, the identifiability is not strictly binary but depends on the conditioning of $R$, which is influenced by the power of the first $L_A$ coefficients compared to the power of the final $L_{G_{\text{N}}}-L_A$ coefficients of the forward path feedforward filter. This effect is apparent from $\kappa(R)$ in Fig. \ref{fig:FIR_cond}, which is smoothly varying, rather than binary. By increasing the forward path delay $\alpha$ or making $\hat{F}_0(q)$ time-varying, $\kappa(R)$ improves, resulting in an increased ASG. Additionally, when $L_{G_{\text{N}}}$ is large, the curves for different $\alpha$'s converge to one another as the coefficients exceeding $L_A$ become dominant in power.

In Fig. \ref{fig:FIR_comp}, the forward path $G_{\text{FIR}}(q)$ is compared to the forward paths $G_{\text{Delay1}}(q)$ and $G_{\text{Delay2}}(q)$ when $\alpha=1$. In real-life speech $G_{\text{Delay2}}(q)$ attains a larger ASG than $G_{\text{Delay1}}(q)$ due to the larger delay in $G_{\text{Delay2}}(q)$ decorrelating the microphone and loudspeaker signals to a larger extent. While using $G_{\text{FIR}}(q)$ does not attain the same performance for real-life speech as the equally long filter $G_{\text{Delay2}}(q)$, governed by the fact that making the first several coefficients of $G_{\text{FIR}}(q)$ non-zero, (versus setting these to zero in $G_{\text{Delay2}}(q)$)  indeed increases $\kappa(R)$, it does outperform $G_{\text{Delay1}}(q)$, still attains identifiability, and allows for a more flexible design of the forward path. Furthermore, $G_{\text{FIR}}(q)$ can outperform $G_{\text{Delay1}}(q)$ through the use of additional filter coefficients, which then counteract the ill-conditioning imposed by the first $L_A$ non-zero coefficients.

\subsection{NOISE-INFLUENCE} \label{subsec:noise_influence}
\begin{figure}
	\centering
	\includegraphics[width=0.5\linewidth]{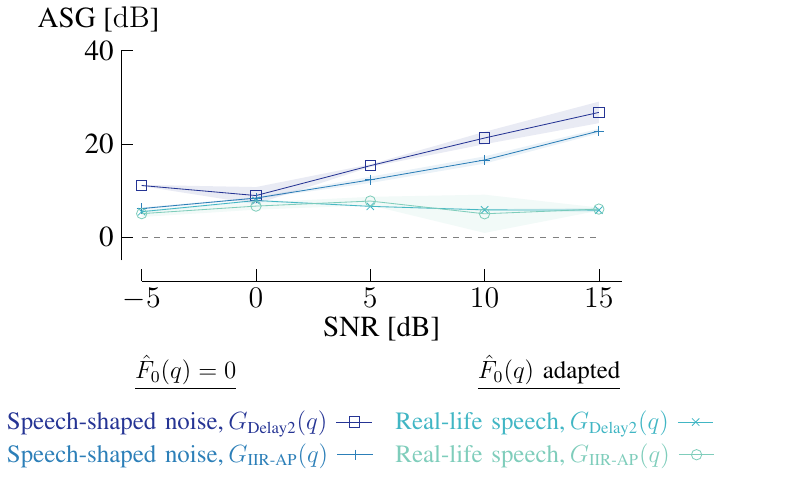}
	\caption{Added stable gain (ASG) for $G_{\text{Delay2}}(q)$ and $G_{\text{IIR-AP}}(q)$ with $\alpha=1$, $L_{G_{\text{N}}}=15$, and $L_D=L_A=10$ as a function of the signal-to-noise ratio (SNR) for speech-shaped noise with $\hat{F}_0(q)=0$, and for real-life speech with $\hat{F}_0(q)$ adapted at each discrete-time instant. A positive ASG is observed across the entire SNR range.}
	\label{fig:SNR}
\end{figure}

Fig. \ref{fig:SNR} shows the ASG for $G_{\text{Delay2}}(q)$ and $G_{\text{IIR-AP}}(q)$ as a function of the SNR of the input signal, with $\alpha=1$, $L_{G_{\text{N}}}=15$, and $L_D=L_A=10$. Two scenarios are considered, one with speech-shaped noise with no $\hat{F}_0(q)$ inserted, i.e., $\hat{F}_0(q)=0$, and  one with real-life speech where $\hat{F}_0(q)$ is replaced with $\hat{F}(q)$ at each discrete-time instant. For the scenario with speech-shaped noise, the ASG is considerably affected by the SNR as the ground-truth was generated according to an AR model, of which the result is perturbed by the babble noise. Nevertheless, even at $\text{SNR}=\SI{-5}{\decibel}$, the ASG remains positive showing a degree of robustness of the identification and resulting ASG against noise. As real-life speech cannot be exactly modelled using an AR model, the identification in the real-life speech scenario is less affected by the SNR, thereby showing an essentially near-constant ASG across the SNR range. As this ASG is furthermore positive across the SNR range, robustness against noise is again observed. The same trends are observed both for $G_{\text{Delay2}}(q)$ and $G_{\text{IIR-AP}}(q)$.

\section{CONCLUSION} \label{sec:conclusion}
In this paper, the identifiability of one specific prediction error method (PEM) feedback cancellation algorithm, the two-channel adaptive feedback canceller (2ch-AFC), has been studied. The delay-based condition stating that identifiability is obtained when the delay in the forward path is equal to or exceeds the order of the autoregressive (AR) model of the input signal is generalised to an invertibility-based condition stating that identifiability is obtained when the order of the forward path feedforward filter exceeds the order of the AR model. The delay-based condition can be seen as a special case of the invertibility-based condition when the coefficients up to the order of the AR model are set equal to zero. The proof is nevertheless specifically tailored to the 2ch-AFC algorithm, rather than being applicable to the whole class of PEM-based feedback cancellation algorithms. 

Additionally, this condition has been verified through simulations. Although in practice the identifiability also seems to depend on the power of the first AR model order coefficients of the forward path feedforward filter with respect to the power of the other coefficients. Indeed, under these circumstances the correlation matrix of which the inverse is used in this 2ch-AFC algorithm can be ill-conditioned, and its condition number can serve as a measure for identifiability.
\clearpage
\bibliographystyle{IEEEtran} 
\bibliography{ref}
\end{document}